\documentstyle[12pt]{article}
\makeatletter
\newdimen\normalarrayskip              
\newdimen\minarrayskip                 
\normalarrayskip\baselineskip
\minarrayskip\jot
\newif\ifold             \oldtrue            
\def\arraymode{\ifold\relax\else\displaystyle\fi} 
\def\eqnumphantom{\phantom{(\theequation)}}     
\def\@arrayskip{\ifold\baselineskip\z@\lineskip\z@
     \else
     \baselineskip\minarrayskip\lineskip2\minarrayskip\fi}
\def\@arrayclassz{\ifcase \@lastchclass \@acolampacol \or
\@ampacol \or \or \or \@addamp \or
   \@acolampacol \or \@firstampfalse \@acol \fi
\edef\@preamble{\@preamble
  \ifcase \@chnum
     \hfil$\relax\arraymode\@sharp$\hfil
     \or $\relax\arraymode\@sharp$\hfil
     \or \hfil$\relax\arraymode\@sharp$\fi}}
\def\@array[#1]#2{\setbox\@arstrutbox=\hbox{\vrule
     height\arraystretch \ht\strutbox
     depth\arraystretch \dp\strutbox
     width\z@}\@mkpream{#2}\edef\@preamble{\halign \noexpand\@halignto
\bgroup \tabskip\z@ \@arstrut \@preamble \tabskip\z@ \cr}%
\let\@startpbox\@@startpbox \let\@endpbox\@@endpbox
  \if #1t\vtop \else \if#1b\vbox \else \vcenter \fi\fi
  \bgroup \let\par\relax
  \let\@sharp##\let\protect\relax
  \@arrayskip\@preamble}
\def\eqnarray{\stepcounter{equation}%
              \let\@currentlabel=\theequation
              \global\@eqnswtrue
              \global\@eqcnt\z@
              \tabskip\@centering
              \let\\=\@eqncr
              $$%
 \halign to \displaywidth\bgroup
    \eqnumphantom\@eqnsel\hskip\@centering
    $\displaystyle \tabskip\z@ {##}$%
    &\global\@eqcnt\@ne \hskip 2\arraycolsep
         $\displaystyle\arraymode{##}$\hfil
    &\global\@eqcnt\tw@ \hskip 2\arraycolsep
         $\displaystyle\tabskip\z@{##}$\hfil
         \tabskip\@centering
    &{##}\tabskip\z@\cr}
\makeatother

\def\beq{\begin{equation}}
\def\eeq{\end{equation}}
\def\bea{\begin{eqnarray}}
\def\eea{\end{eqnarray}}

\def\Bf#1{\mbox{\boldmath $#1$}}

\def\balpha{{\Bf\alpha}}
\def\bbeta{{\Bf\beta}}
\def\bnu{{\Bf\nu}}
\def\bmu{{\Bf\mu}}
\def\bphi{{\Bf\phi}}
\def\bN{{\Bf N}}

\def\bJ{{\Bf J}}
\def\be{{\Bf e}}

\def\bsN{\hbox{${\small {\Bf N}}$}}

\def\bsalpha{\hbox{${\small {\Bf\alpha}}$}}
 \def\bsbeta{\hbox{${\small  {\Bf\beta}}$}}
\def\bsphi{{\Bf\phi}}

\def\W{{\rm W}}

\def\w{$W$-algebra }
\def\wi{$W$-invariance }
\def\mmms{multi-matrix models }
\def\mmm{multi-matrix model }
\def\nn{\nonumber}
\def\mm{matrix model }
\def\mms{matrix models }

\begin{document}

\begin{titlepage}
\begin{center}
{{\it P.N.Lebedev Institute preprint} \hfill FIAN/TD-5/92\\
{\it I.E.Tamm Theory Department} \hfill hepth/9209100\\
\begin{flushright}{May 1992}\end{flushright}
\vspace{0.1in}{\Large\bf On the Continuum Limit\\
of the Conformal Matrix Models\footnote{
Talk presented at the Conference ``Modern Problems in Quantum
Field Theory, Strings, Quantum  Gravity'',
Kiev, 8-17 June 1992}}\\[.4in]
{\large   A. Mironov}\\
\bigskip {\it  P.N.Lebedev Physical
Institute \\ Leninsky prospect, 53, Moscow, 117 924, Russia},
\footnote{E-mail address: tdparticle@glas.apc.org   \&
mironov@sci.fian.msk.su}\\ \smallskip
\bigskip {\large S. Pakuliak}\\
 \bigskip {\it Institute for Theoretical Physics\\
 Kiev 252130, Ukraine}
\footnote{E-mail address: spakuliak@glas.apc.org}
\footnote{Address after July 1992, RIMS, Kyoto University, Kyoto-606,
e-mail:  pakuliak@kurims.kyoto-u.ac.jp}} \end{center}
\bigskip \bigskip
\begin{abstract}
The double scaling limit of a  new class of the \mmms proposed in \cite{MMM91},
which possess the
$W$-symmetry at the
discrete level, is investigated in details.
These models are demonstrated to
fall into the same universality class as the standard \mmms. In particular, the
transformation of the \w at the discrete level into the continuum one of the
paper
\cite{FKN91a} is proposed, the corresponding partition functions being
compared. All calculations are demonstrated in full in the first non-trivial
case of $W^{(3)}$-constraints.
\end{abstract}

\end{titlepage}

\newpage
\setcounter{page}{2}
\setcounter{footnote}{0}
\section{Introduction}

The idea of the description of 2$d$ theories of conformal matter coupled to
gravity
through the simpler models of the lattice type, which fall into the same
universality classes,
suggested in \cite{Kaz85,Dav85} led to an explosure of the interest to
different
matrix models in 1989, when it was understood that the most convenient choice
of the lattice models is related to the triangulations of random surfaces
(in spirit of Regge approach \cite{Regge}), which can be coded in terms of
\mms.
More concretely, there was proposed
a class of \mmms whose (multi-)critical points correspond to minimal
series of conformal matter coupled to gravity
\cite{Kaz89,BK90,DS90,GM90a,GM90b,BDKS90,CGM90,GM90c}. The general description
of the (double scaling) continuum limit of these models
essentially involves the following two properties of the double scaled
partition function: it should be the $\tau$-function of proper reduced
KP hierarchy and it should satisfy the \w \cite{Dou90,FKN91a,DVV91a}. The
combination
of these two properties leads to the string equation.

Indeed, it was  the string equation that was derived firstly, and the general
properties of the \mms were guessed later from the properties of the string
equation. Unfortunately, it is very difficult to prove them completely, as the
the structure of the theory is very complicated when considering
higher multi-critical points and/or \mmms.
{}From the other hand, let us point out that the determing properties of the
\mmm  partition function are observed {\it after} taking the continuum limit.
Therefore, it is reasonable to find out new \mmms with these properties
 at the {\it discrete} level.
In fact, at the discrete level one usually has a property of
manifest distinguishing
 the integrability
and the \wi simultaneously. Due to this fact, one can
effectively
investigate the continuum limit by observing {\it only} \w (by modulo one
subtlety,
see below the subsect.5.2). As for the integrability, it is the standard thing
that one can
continue the model off critical point preserving this property. This
continuation is rather natural and simultaneously gives rise to a proper
regularization of the continuum theory (we also know this phenomenon in quantum
integrable systems, where to regularize the theory one can put it on the
lattice preserving the quantum integrability - see, for example,
\cite{Korepin}).

All this was partially realized previously. Namely, there was proved
that at the discrete level the partition function of the Hermitian one-\mm is
the $\tau$-function of Toda chain hierarchy \cite{GMMMO91} and satisfies the
Virasoro constraints \cite{MM90,AJM90}. It was also shown \cite{MMMM91}
that these
constraints turn into the continuum constraints of the paper \cite{FKN91a}.
Moreover,
one can check \cite{GMMMMO91} that in this continuum limit Toda chain hierarchy
really turns into KdV hierarchy which corresponds to the double scaling limit
of
the Hermitian one-\mm \cite{Dou90,FKN91a,DVV91a}.

Unfortunately, the situation is not so clear in the \mmm case. In the
general description the standard \mmms
proposed in \cite{BDKS90,CGM90,GM90c}
have no any symmetry at the discrete level (though they
still correspond to an integrable system, that is, to Toda lattice hierarchy
\cite{GMMMO91,KMMOZ91,KMMM92}). It does the proof of their \wi in the
continuum limit very difficult. There were also suggested
other discrete \mmms with the same integrability property
and \wi \cite{GvN91,MMM92b}. Their continuum limit was
investigated \cite{GvN91}, but the complete answer is not obtained up to now
due
to technical difficulties.

At last, in the paper \cite{MMM91} a rather natural \mmm was introduced, with
the property of \wi included by definition. This model has less
trivial integrable structure and corresponds to multi-component KP hierarchy
with a special reduction of AKNS type \cite{KMMMP92}.
This model will be the main subject under consideration in the
paper.
As the corresponding partition function  can be naturally presented in the form
of the
correlator in conformal theory, we call it "conformal \mmm" (CMM).

We consider the paper as one of the series of two papers devoted to the
investigation of CMM.
Another one \cite{KMMMP92} mostly discusses the the integrability, while
the continuum limit
is described in less details. The present paper thoroughly deals with the
continuum
limit of CMM. But, for the sake of completeness, we describe
general properties of the discrete model. More concretely, in the section 2
the definition
of CMM is introduced, and some of its general properties are discussed.
Here we are interested only in the continuum limit of \w,
the continuum limit of arbitrary multicomponent KP hierarhy of such type
being out of
the scope of this paper.
In the sect.3 the general approach to taking the continuum limit is discussed.
It is considered for the simplest case of the Virasoro constraints
in the one-\mm in
rather invariant terms \cite{MMMM91}, and some natural generalizations
are
proposed. It turns out that this invariant formulation naturally
continues to the general case due to the very special form of the
discrete \w
resembling the structure of the continuum \w of the paper \cite{FKN91a}
(indeed, just the "conformal" form of the partition function in this model is
in
charge of this). In the sect.4 the simplest non-trivial case of
two-\mm
($W^{\{3\}}$-constraints) is considered in full details, and the general
case is discussed in the sect.5,
some important but tedious calculations being
shifted to the Appendix. The conclusion contains the list of the main
results
as well as some general discussion.

\section{Conformal \mmms}
\setcounter{equation}{0}
\subsection{Formulation of the model}

To begin with, we would like to introduce conformal \mmms in accordance with
the paper \cite{MMM91} (see also \cite{KMMMP92}).
First, we show that the simplest example of discrete Hermitian one-matrix
model can be easily reformulated in these terms.

Indeed, Hermitian one-matrix model $(p=2)$ can be defined as a solution to
discrete Virasoro constraints:
$$
L_nZ_{2,N}[t] = 0,\ \  \   n \geq  -1
$$
\beq\label{3}
L_n \equiv \sum ^\infty _{k=0}kt_k\partial /\partial t_{k+n} +
\sum _{a+b=n}\partial ^2/\partial t_a\partial t_b
\eeq
$$
\partial Z_{2,N}/\partial t_0 = -NZ_{2,N}
$$
The Virasoro generators (\ref{3})
 have the well-known form of the Virasoro operators in
the theory of one free scalar field. If we look for such solution in terms of
holomorphic components of the scalar field
\bea\label{4}
\phi (z) &=  \hat q + \hat p \log z  + \sum _{k\neq 0} {J_{-k}\over k}
z^{-k}\nn\\
\  [J_n,J_m] &= n\delta _{n+m,0},  \ \ \     [\hat q,\hat p] = 1
\eea
the procedure is as follows. Define vacuum states
\bea\label{5}
J_k|0\rangle  &= 0, \ \ \  \langle N|J_{-k} = 0, \ \ \    k > 0\nn\\
\hat p|0\rangle  &= 0, \ \ \   \langle N|\hat p = N\langle N|,
\eea
the stress-tensor
\beq\label{6}
T(z) = {1\over 2}{:}[\partial \phi (z)]^2{:} = \sum    T_nz^{-n-2},\quad
T_n = {1\over 2}\sum _{k>0}{:}J_{-k}J_{k+n}{:} +
{1\over 2}\sum _{{a+b=n}\atop{a,b\geq 0}}J_aJ_b,
\eeq
\beq\label{6a}
 T_n|0\rangle  = 0,  \ \ \    n \geq  -1
\eeq
and the Hamiltonian
\bea\label{7}
H(t) &= {1\over \sqrt{2}} \sum _{k>0}t_kJ_k =
\oint_{C_0}V(z)j(z)\nn\\
V(z) &= \sum _{k>0}t_kz^k, \ \  \   j(z) = {1\over \sqrt{2}}\partial \phi (z).
\eea
Now one can easily construct a ``conformal field theory" solution to
(\ref{3}) in two steps. First,
\beq\label{8}
L_n\langle N|e^{H(t)}\ldots = \langle N|e^{H(t)}T_n\ldots
\eeq
can be checked explicitly. As an immediate consequence, any correlator of the
form
\beq\label{9}
\langle N|e^{H(t)}G|0\rangle
\eeq
($N$  counts the number of zero modes of  $G$) gives a solution to (\ref{3})
provided
\beq\label{10}
[T_n,G] = 0, \ \ \  n \geq  -1.
\eeq
Second, the conformal solution to (\ref{10})
(and therefore to (\ref{3})) comes from the
properties of $2d$ conformal algebra. Indeed, any solution to
\beq\label{11}
[T(z),G] = 0
\eeq
is a solution to (\ref{10}),
and it is well-known that the solution to (\ref{11}) is a
function of {\it screening charges}
\beq\label{12}
Q_\pm  = \oint J_\pm  = \oint
e^{\pm \sqrt{2}\phi }.
\eeq
With a selection rule on zero mode it gives
\beq\label{13}
G = \exp \ Q_+ \rightarrow  {1\over N!}Q^N_+
\eeq
Of course, the general case must be  $G \sim  Q^{N+M}_+Q^M_-$. Nevertheless,
this choice of one of the two possible screening operators has a clear
algebraic sense which we will discuss below in this section.
It can be justified  by the special
prescription for integration contours, proposed in \cite{MMM91},
which implies that the
dependence of $M$ can be irrelevant and one can just put  $M = 0$).
In this case the solution
\beq\label{14}
Z_{2,N}[t] = \langle N|e^{H(t)}\exp Q_+|0\rangle
\eeq
after computation of the free theory correlator
gives
well-known result
\bea\label{15}
Z_{2,N} &= (N!)^{-1}\int   \prod ^N_{i=1}dz_i \exp  \left( - \sum
t_kz^k_i\right)  \Delta ^2_N(z) =\nn\\
&= (N!{\rm Vol}\ U(N))^{-1}\int   DM\ \exp \left( - \sum    t_kM^k\right)
\\
\Delta _N &= \prod ^N_{i<j}(z_i - z_j)\nn
\eea
in the form of multiple integral over spectral parameters or integration over
Hermitian matrices.

In the case of  $p=2$  (Virasoro) constraints this is just a useful
reformulation of the Hermitian 1-matrix model. However, in what follows we are
going to use this point of view as a constructive one. Indeed, instead of
considering a special direct multi-matrix generalization of (\ref{15})
\cite{BDKS90,CGM90,GM90c} one
can use powerful tools of conformal theories, where it is well known how to
generalize almost all the steps of above construction: first, instead of
looking for a solution to Virasoro constraints one can impose extended
Virasoro or  $W$-constraints on the partition function. In such case one would
get Hamiltonians in terms of {\it multi}-scalar field theory, and the second
step is generalized directly using {\it screening charges} for  $W$-algebras.
The general scheme looks as follows

(i)  Consider Hamiltonian as a linear combination of the Cartan currents of a
level one Kac-Moody algebra  ${\cal G}$
\beq\label{16}
H(t^{(1)},\ldots,t^{({\rm rank}\ {\cal G})}) =
\sum _{\lambda ,k>0}t^{(\lambda )}_k\bmu _\lambda \bJ_k,
\eeq
where $\{\bmu_i\}$ are basis vectors in Cartan hyperplane, which,
say for $SL(p)$ case are chosen to satisfy
$$
\bmu_i\cdot \bmu_j=\delta_{ij}-{1\over p},  \ \ \ \sum_{j=1}^p
\bmu_j=0.
$$

(ii)  The action of differential operators  $W^{(a)}_i$ with respect to
times  $\{t^{(\lambda )}_k\}$ can be now defined from the relation
\beq\label{17}
W^{(a)}_i\langle \bN|e^{H(\{t\})}\ldots =
\langle \bN|e^{H(\{t\})}{\rm W}^{(a)}_i\ldots\ , \ \ \   a=2,\ldots,p;  \ \ \
i\geq 1-a,
\eeq
where
\bea\label{18}
\W^{(a)}_i &= \oint z^{a+i-1}\W^{(a)}(z)\nn\\
\W^{(a)}(z) &= \sum  _\lambda  {:}[\bmu _\lambda \partial \bphi (z)]^a{:}
 + \ldots
\eea
are  spin-$a$ W-generators of  $\W_p$-algebra written in terms of
rank$\;{\cal G}$-component scalar fields \cite{FL88}.

(iii)  The conformal solution to $W$-constraints arises in the form
\beq\label{19}
Z^{\rm CMM}_{p,\bsN}[\{t\}] = \langle \bN|e^{H(\{t\})}G\{Q
^{(\alpha)} \}|0\rangle
\eeq
where  $G$  is an exponential function of screenings of level one Kac-Moody
algebra
\beq\label{20}
Q^{(\alpha)}  = \oint J^{(\alpha)}  = \oint e^{\bsalpha \bsphi }
\eeq
$\{\balpha \}$ being roots of finite-dimensional simply laced
Lie algebra ${\cal G}$. (For the case of non-simply
laced case see \cite{FF92}. Below ${\cal G}=SL(p)$ if not stated otherwise.)
The correlator (\ref{19}) is still a free-field correlator
and the computation gives it again in a multiple integral form
\bea\label{21}
Z^{\rm CMM}_{p,\bsN}[\{t\}] &\sim  \int   \prod  _\alpha
\left[ \prod ^{N_\alpha }_{i=1}dz^{(\alpha )}_i \exp \left( -
\sum _{\lambda ,k>0}t^{(\lambda )}_k(\bmu _\lambda \balpha )(z^{(\alpha )}_i)^k
\right) \right] \times \nn\\
&\times \prod _{(\alpha ,\beta )}\prod ^{N_\alpha }_{i=1}
\prod ^{N_\beta }_{j=1}(z^{(\alpha )}_i- z^{(\beta )}_j)^{\bsalpha \bsbeta }
\eea
The expression (\ref{21})
is what we shall study in this paper: namely the solution to
discrete  $W$-constraints
which can be written as multiple integral over
spectral parameters  $\{z^{(\alpha )}_i\}$ (this integral is sometimes called
``eigenvalue model"). The difference with the one-matrix case (\ref{15})
 is that the
expressions (\ref{21}) have rather complicated representation in terms of
multi-matrix integrals. Namely, the only non-trivial (Van-der-Monde) factor can
be rewritten in the (invariant) matrix form:
\bea\label{21a}
\prod ^{N_\alpha }_{i=1} \prod ^{N_\beta }_{j=1}(z^{(\alpha )}_i-
z^{(\beta )}_j)^{\bsalpha \bbeta } = \left[ \det \{M^{(\alpha )}\otimes I -
I\otimes M^{(\beta )}\}\right] ^{\bsalpha \bsbeta },
\eea
where $I$ is the unit matrix. Still this is a model with a chain of matrices
and with closest neibour interactions only (in the case of $SL(p)$).

Now we would like to say some words on the general structure of this model.
Let us point out that its partition function is nothing but the correlation
function of objects which have clear algebraic meaning. Indeed,
the time dependent exponential is generated by Cartan currents of $SL(p)$,
and the exponentials of the screening charges correspond to the exponentials
of other (non-Cartan)
generators of $SL(p)$. But we should stress that again it should not
be an arbitrary combination of these generators, but only of any $p-1$ of
them. Indeed, the different choices of these $p-1$ generators correspond to the
{\it different} models but with the same properties of the integrability and
the \wi.

In fact, it is very immediate thing to fermionize these expressions and
merely write down the expression for proper $\tau$-function of reduced KP
hierarchy of
the generalized
AKNS type \cite{AKNS74,UT84}
in the form of fermionic correlator \cite{KMMMP92}.
This generalized AKNS reduction has its origin in the reducing
from the general case of $GL(p)$ algebra to its simple subalgebra $SL(p)$
\cite{KMMMP92}, when considering proper
$W^{(p)}$-algebra \cite{FL88}\footnote{Indeed,
all this has an interpretation immediately in terms of proper Hamiltonian
reduction - see also \cite{MMM91,GMM90}.}.

The purpose of this paper is to show that CMM defined by (\ref{21}) as a
solution to the $W$-constraints
possesses a natural continuum limit. To pay for these advantages one
should accept a slightly less elegant matrix integral with the entries like
 (\ref{21a}).

\subsection{On the proper basis for CMM}

Now we would like to discuss briefly the manifest expressions for constraint
algebras in terms of time variables.

The first non-trivial example (which we use as a demonstrating example
in the section 4)
is the $p=3$ associated with
Zamolodchikov's  $W_3$-algebra \cite{Zam85}
and serves as alternative to 2-matrix model.
In this particular case one obtains
\beq\label{22}
H(t,\bar t) = {1\over \sqrt{2}} \sum _{k\geq 0}(t_kJ_k + \bar t_k\bar J_k)
\eeq
\bea\label{23}
\W^{(2)}_n = &L_n = \sum ^\infty _{k=0}(kt_k\partial /\partial t_{k+n} +
k\bar t_k\partial /\partial \bar t_{k+n}) +\nn\\
&+ \sum _{a+b=n}(\partial ^2/\partial t_a\partial t_b+
\partial ^2/\partial \bar t_a\partial \bar t_b)
\eea
\bea\label{24}
\W^{(3)}_n = &\sum _{k,l>0}(kt_klt_l\partial /\partial t_{k+n+l} -
k\bar t_kl\bar t_l\partial /\partial t_{k+n+l}
-2kt_kl\bar t_l\partial /\partial \bar t_{k+n+l})+\nn\\
&+ 2\sum  _{k>0}\left[ \sum _{a+b=n+k}(kt_k\partial ^2/\partial t_a
\partial t_b - kt_k\partial ^2/\partial \bar t_a\partial \bar t_b -
2k\bar t_k\partial ^2/\partial t_a\partial \bar t_b)\right]  +\nn\\
&+ {4\over 3}\sum _{a+b+c=n}(\partial ^3/\partial t_a\partial t_b\partial t_c-
\partial ^3/\partial t_a\partial \bar t_b\partial \bar t_c),
\eea
where times  $t_k$ and  $\bar t_k$ correspond to the two orthogonal directions
in  $SL(3)$  Cartan plane.
(We use the standard specification of the Cartan
basis:  $\be = \balpha _1/\sqrt{2}$, $\bar\be =
\sqrt{3}\bnu _2/\sqrt{2}$.)
In this case
one has six screening charges  $Q^{(\pm \alpha _i)}$ $(i = 1,2,3)$
which commute with
\beq\label{25}
\W^{(2)}(z) = T(z) = {1\over 2}{:}[\partial \bphi (z)]^2{:}
\eeq
and
\beq\label{26}
\W^{(3)}(z) = \sum ^3_{\lambda =1}{:}(\bmu _\lambda \partial \bphi (z))^3{:}\
\ , \eeq
where  $\bmu _\lambda $ are vectors of one of the fundamental representations
({\bf 3} or $\bar {\bf 3})$ of  $SL(3)$.

This basis was originally used in \cite{MMM91} as it just corresponds to
the integrable flows, for the continuum limit we will use another basis in the
Cartan plane connected with $t \pm \bar t$.
(In other words, this is the question what is the
proper reduction, or what combinations of the ``integrable" times should be
eliminated.)

To begin with, we consider the simplest non-trivial case of $p=3$. Then
introducing the scalar fields
\beq\label{d1}
\partial \phi ^{(1)}(z) = \sum  _k kt^{(1)}_kz^{k-1} + \sum  _k
{\partial \over \partial t^{(2)}_k} z^{-k-1},
\eeq
\beq\label{d2}
\partial \phi ^{(2)}(z) = \sum  _k kt^{(2)}_kz^{k-1} + \sum  _k
{\partial \over \partial t^{(1)}_k} z^{-k-1},
\eeq
with  $t^{(1)}_k= (i\bar t_k+t_k)/2\sqrt2$,
$t^{(2)}_k= (i\bar t_k-t_k)/2\sqrt2$, one
obtains the expressions:
\beq\label{d3}
W^{(2)}(z) = {1\over2}{:}\partial \phi ^{(1)}(z)\partial \phi ^{(2)}(z){:}\ \ ,
\eeq
\beq\label{d4}
W^{(3)}(z) = {1\over3\sqrt3}\sum  _i {:}(\partial \phi ^{(i)}(z))^3{:}\ \ .
\eeq
instead of (\ref{23}) and (\ref{24}).

This choice of basis in the Cartan plane is adequate to the continuum limit of
the system under consideration, as the latter one is described by completely
analogous expressions \cite{FKN91a}.
Now let us describe this basis in more invariant
terms and find the generalization to arbitrary $p.$

Comparing (\ref{d4}) with (\ref{26}), we can conclude that  $\partial
\phi ^{(i)}\equiv
\bbeta _i\partial \bphi $  corresponds to the basis
\beq\label{d5}
\bbeta _{1,2} = {1\over 2}(\sqrt{3}\bmu _2 \pm  i\balpha _2).
\eeq
This basis has the properties
\beq\label{d6}
\bbeta _1\cdot \bbeta _2 = 1,  \ \ \  \bbeta _1\cdot \bbeta _1 = 0,  \ \ \
\bbeta _2\cdot \bbeta _2 = 0.
\eeq

Now it is rather evident how this basis should look in the case of
general $p$. Due to \cite{FKN91a} we can guess what is the choice of the proper
scalar fields:
\beq\label{d7}
\partial \phi ^{(i)}(z) = \sum  _k kt^{(i)}_kz^{k-1} + \sum  _k
{\partial \over \partial t^{(p-i)}_k} z^{-k-1}.
\eeq
This choice certainly corresponds to the basis with defining property (it can
be observed immediately from the relations (\ref{d7}) and (\ref{17})):
\beq\label{d8}
\bbeta _i\cdot \bbeta _j = \delta _{p,i+j},
\eeq
the proper choice of the Hamiltonians in (\ref{16}) being
\beq\label{d9}
H = \sum _{i,k}t^{(i)}_k\bbeta _i\cdot \bJ_k,
\eeq
what determines new times adequate to the continuum limit.

Let us construct the basis (\ref{d8}) in a manifest way.
To begin with, we define a set of vectors $\{\mu _i\}$ with the property:
\beq\label{d10}
\bmu _i\cdot \bmu _j = \delta _{ij} - {1\over p}, \ \ \  \sum  _i \bmu _i = 0.
\eeq
The $W^{(n)}$-algebra can be written in this basis as follows \cite{FL88}:
\beq\label{d11}
W^{(n)} = (-)^{n+1} \sum _{1\leq j_1<\ldots<j_k\leq p}~ \prod ^n_{m=1}
{:}(\bmu _{j_m}\cdot \partial \bphi ){:}\ ,  \ \ \   n=2,\ldots,p.
\eeq
Now the basis (\ref{d8}) can be constructed from (\ref{d10})
by diagonalization of the
following cyclic permutation \cite{FKN91a,FKN91b}:
\beq\label{d12}
\bmu _i \rightarrow  \bmu _{i+1}, \ \ \  \bmu _p \rightarrow  \bmu _1\ \ \
i=1,\ldots,p-1 .
\eeq
This transformation has $\{\bbeta _i\}$ as its eigenvectors, their manifest
expressions being of the form:
\beq\label{d13}
\bbeta _k = {1\over \sqrt p} \sum ^p_{j=1} \exp \{{2\pi i\over p}jk\}
\bmu _j,  \ \ \   k=1,2,\ldots,p-1.
\eeq
It is trivial to check that the properties (\ref{d8}) are indeed satisfied.
One can immediately rewrite the corresponding $W$-generators in the basis of
$\bbeta _i$'s. After all, one obtain the expressions similar to the continuum
$W$-generators \cite{FKN91a,FKN91b},
but with the scalar fields defined as in (\ref{d7})
and without
the ``anomaly" corrections appearing in the continuum case due to the twisted
boundary conditions. These corrections can be correctly reproduced by taking
the $p$-th root of the partition function as well as simultaneously doing the
reduction (see the sects.4-5).

Thus, the proposed procedure allows one to take the continuum limit immediately
transforming the scalar fields as elementary building blocks. Neveretheless,
before immediate doing of the continuum limit for any CMM let us consider
this in the simplest case of the Hermitian one-\mm \cite{MMMM91} and
get some insight for the general case.

\section{Double-scaling limit of CMM: preliminary comments}
\subsection{Results of [20] for the one-\mm}

To begin with let us briefly remind the main points of \cite{MMMM91}.

It has been suggested in \cite{FKN91a} that the square root of the
partition function of the continuum limit of
one-\mm is subjected to the Virasoro constraints
\beq\label{b1}
{\cal L}_n^{\rm cont}\sqrt{Z^{\rm ds}}=0,\quad n\geq -1,
\eeq
where
\bea \label{b2}
{\cal L}^{\rm cont}_n=&\sum_{k=0}\left(k+{1\over2}\right)
T_{2k+1}{\partial\over\partial T_{2(k+n)+1}}+G
\sum_{0\leq k\leq n-1}{\partial^2\over\partial T_{2k+1}\partial
T_{2(n-k-1)+1}}+\nn\\
+&{\delta_{0,n}\over16} +{\delta_{-1,n}T_1^2\over(16G)}
\eea
are modes of the stress tensor
\beq\label{b3}
{\cal T}(z)= {1\over 2}{:}(\partial\Phi(z))^2{:} - {1\over16z^2}
=\sum{{\cal L}_n\over z^{n+2}},
\eeq
where
\beq\label{b3a}
\partial\Phi(z)=\sum_{n\geq0}\left(\left(n+{1\over2}\right)
T_{2n+1}z^{n-{1\over2}}+{\partial\over\partial T_{2n+1}}
z^{-n-{3\over2}}\right).
\eeq

It was shown in  \cite{MMMM91} that these equations which reflect the
$W^{(2)}$-invariance of the partition function of the continuum model can
be deduced from analogous constraints in Hermitian one-\mm by taking the
double-scaling continuum limit.
The procedure (generalized below to CMM) is as follows.

The partition function of Hermitian one-\mm  can be written in the form
\beq	\label{b4}
Z\{t_{k}\}=\int{\cal D}M\exp{\rm Tr}\sum_{k=0}t_{k}
M^{k}
\eeq
and satisfies \cite{MM90,AJM90} the discrete Virasoro constraints  (\ref{3})
\bea  \label{b5}
&L^{\rm H}_{n}Z=0,\quad n\geq0 \nn\\
&L^{\rm H}_{n}=\sum_{k=0}kt_{k}{\partial\over\partial t_{k+n}}+
\sum_{0\leq k\leq n}{\partial ^2\over \partial t_{k}\partial t_{n-k}}.
\eea
In order to obtain the above-mentioned relation between $W$-invariance of the
discrete and continuum models one has to consider a reduction of  model
(\ref{b4}) to the pure even potential $t_{2k+1}=0$.

Let us denote by the $\tau_N^{\rm red}$ the partition function of the
reduced \mm
\beq	\label{a1}
\tau^{\rm red}_{N}\{t_{2k}\}=\int{\cal D}M\exp{\rm Tr}\sum_{k=0}t_{2k}
M^{2k}
\eeq
and consider the following change of the time variables
\beq\label{a5}
g_m=\sum_{n\geq m}{(-)^{n-m}\Gamma\left(n+{3\over2}\right)
a^{-n-{1\over2}}\over(n-m)!\Gamma\left(m+{1\over2}\right)}T_{2n+1},
\eeq
where $g_m \equiv mt_{2m}$ and this expression can be used also for the zero
discrete time $g_0 \equiv N$ that plays the role of the dimension of matrices
in the one-matrix model. Derivatives with respect to  $t_{2k}$ transform as
\beq\label{a6}
{\partial\over\partial t_{2k}}=\sum_{n=0
}^{k-1}{\Gamma\left(k+{1\over2}\right)
a^{n+{1\over2}}\over(k-n-1)!\Gamma\left(n+{3\over2}\right)}{\partial
\over\partial \tilde T_{2n+1}},
\eeq
where the auxiliary continuum times $\tilde T_{2n+1}$ are connected with
``true'' Kazakov continuum times $T_{2n+1}$ via
\beq\label{a7}
T_{2k+1}=\tilde T_{2k+1}+a{k\over k+1/2}\tilde T_{2(k-1)+1},
\eeq
and coincide with $T_{2n+1}$ in the double-scaling limit when $a\to0$.

Let us rescale the partition function of the reduced one-\mm by
exponent of quadratic form of the auxiliary times $\tilde T_{2n+1}$
\beq\label{a11}
\tilde\tau=\exp\left(-{1\over2}\sum_{m,n\geq0}A_{mn}\tilde T_{2m+1}
\tilde T_{2n+1} \right)\tau^{\rm red}_N
\eeq
with
\beq\label{a12}
A_{nm}={\Gamma\left(n+{3\over2}\right)\Gamma\left(m+{3\over2}\right)\over
2\Gamma^2\left({1\over2}\right)}
{(-)^{n+m}a^{-n-m-1}\over n!m!(n+m+1)(n+m+2)}.
\eeq
Then a direct though tedious calculation \cite{MMMM91} demonstrates that
the relation
\beq\label{a15}
{\tilde{\cal L}_n\tilde\tau\over\tilde\tau}
=a^{-n}\sum_{p=0}^{n+1}C^p_{n+1}(-1)^{n+1-p}
{L_{2p}^{\rm red}\tau^{\rm red}\over \tau^{\rm red}},
\eeq
is valid, where
\bea
L_{2n}^{\rm red} \equiv \sum_{k=0}kt_{2k}{\partial\over\partial t_{2(k+n)}}+
\sum_{0\leq k\leq n}{\partial ^2\over \partial t_{2k}\partial t_{2(n-k)}}
\eea
and
\bea\label{b6}
\tilde{\cal L}_{-1}=&\sum_{k\geq1}\left(k+{1\over2}\right)T_{2k+1}
{\partial\over\partial \tilde T_{2(k-1)+1}}+
{T_1^2\over16},\nn\\
\tilde{\cal L}_{0}=&\sum_{k\geq0}\left(k+{1\over2}\right)T_{2k+1}
{\partial\over\partial \tilde T_{2k+1}},\nn\\
\tilde{\cal L}_n=&\sum_{k\geq0}\left(k+{1\over2}\right)T_{2k+1}
{\partial\over\partial \tilde T_{2(k+n)+1}} \nn\\
&+\sum_{0\leq k \leq n-1}{\partial\over\partial \tilde T_{2k+1}}
{\partial\over\partial \tilde T_{2(n-k-1)+1}} -{(-)^n\over16a^n},\ \ \
n\geq1.
\eea
Here $C^p_n =\frac{n!}{p!(n-p)!}$ are binomial coefficients.

These Virasoro generators differ from the Virasoro generators
(\ref{b2})  \cite{FKN91a,DVV91a}
by terms which are singular in the limit $a\longrightarrow 0$.
At the same time $L_{2p}^{\rm red}\tau^{\rm red}$ at the r.h.s. of (\ref{a15})
do
not need to vanish, since
\bea\label{15aa}
0 = L_{2p}\tau\biggl\vert_{t_{2k+1}=0} =
L_{2p}^{\rm red}\tau^{\rm red} +
\sum_i {\partial^2\tau\over\partial t_{2i+1}\partial t_{2(n-i-1)+1}}
\biggr\vert_{t_{2k+1}=0}.
\eea
It was shown in \cite{MMMM91} that these two origins of difference between
(\ref{b2}) and (\ref{b6}) actually cancel each other, provided eq.(\ref{a15})
is rewritten in terms of the square root $\sqrt{\tilde\tau}$ rather than
$\tilde\tau$ itself:
\beq\label{a16}
{{\cal L}^{\rm cont}_n\sqrt{\tilde\tau}\over\sqrt{\tilde\tau}}
=a^{-n}\sum_{p=0}^{n+1}C^p_{n+1}(-1)^{n+1-p}\left. {L_{2p}\tau\over\tau}
\right\vert_{t_{2k+1}=0}\left( 1+  O(a) \right).
\eeq

\subsection{Generalization of the Ka\-za\-kov variables}

Now we would like to generalize the procedure proposed in the
sect.3.1 to the case of the general \mmm. However, before doing this, we
shall describe a simpler way to guess proper Kazakov variables as well as
tilded time variables and the connection between discrete and continuum
\w. It can be done already at the level of the "leading" (or
"quasiclassical" in accordance with \cite{MM90}) terms, $i.e.$ those
including only the first derivatives. It is possible to do due to the fact
that
the constraint algebra of CMM
strongly resembles the structure of the continuum ${\cal W}$-algebra [2]
(see (\ref{d3}), (\ref{d4})).

To begin with, we would like to consider the simplest multi-matrix case of
two matrices in details. The general case which is completely analogous
to the two-matrix one
but requires considerably more tedious calulations will be considered in the
Appendix.

The leading terms of the $W^{(3)}$ and
${\cal W}^{(3)}$ generators are given
by the formulas
\beq\label{3.1}
\sqrt3 W^{(3){\rm lead}}_n=\sum_{i=1}^2\sum _{k,l>0}kt^{(i)}_klt^{(i)}_l
{\partial\over \partial t^{(3-i)}_{k+n+l}}, \quad g^{(i)}_k=kt^{(i)}_k,
\eeq
\bea\label{3.2}
\sqrt3 {\cal W}^{(3){\rm lead}}=
&\sum_{k,m\geq0}
\left(k+{1\over3}\right)\left(m+{1\over3}\right)T_{3k+1}T_{3m+1}
{\partial\over\partial\tilde T_{3(k+m+n)+1}}\nn\\
+&\sum_{k,m\geq0}
\left(k+{2\over3}\right)\left(m+{2\over3}\right)T_{3k+2}T_{3m+2}
{\partial\over\partial\tilde T_{3(k+m+n-1)+2}}.
\eea
Now, to connect these two algebras, we have to find out
a change of time variables similar to (\ref{a5}),
(\ref{a6}) which gives the relation between leading terms of the generators
(\ref{3.1}), (\ref{3.2}). This problem can be solved in two steps.
First, we make a reduction of the discrete times
\beq    \label{4.4}
t^{(i)}_{3k+1}=t^{(i)}_{3k+2}=0,\ i=1,2,\ k=0,1,\ldots\ \ .
\eeq
Second, we note that the changing of time variables
\beq\label{4.5}
g^{(i)}_m=\sum_{n\geq m}{(-)^{n-m}\Gamma\left(n+1+{i\over3}\right)
a^{-n-{i\over3}}\over(n-m)!\Gamma\left(m+{i\over3}\right)}T_{3n+i}\
,\  i=1,2,
\eeq
\beq\label{4.6}
{\partial\over\partial t^{(i)}_{3k}}=\sum_{n=0
}^{k-1}{\Gamma\left(k+{i\over3}\right)
a^{n+{i\over3}}\over(k-n-1)!\Gamma\left(n+1+{i\over3}\right)}{\partial
\over\partial \tilde T_{3n+i}},\ i=1,2,
\eeq
where $g^{(i)}_m$ is equal to $mt^{(i)}_{3m}\ $\footnote{Similar to the case
considered in the previous subsection we have to include properly into the
set of the independent discrete variables the sizes of matrices $N_i$. These
are
just new variables of zero indices $g^{(i)}_0$.
In the continuum limit they are related
to the cosmological constant (generally there should be many different
cosmological constants). All this can be treated
completely along the line of \cite{MMMM91}, so
we will not discuss it here.},   and the
auxiliary continuum times $\tilde T_{3n+i}$ are connected with
the Kazakov continuum times $T_{3n+i}$ through
\beq\label{4.7}
T_{3k+i}=\tilde T_{3k+i}+a{k\over k+i/3}\tilde T_{3(k-1)+i}.
\eeq
It gives rise to the following relation
between $W^{(3){\rm lead}}$ and
$\hat {\cal W}^{(3){\rm lead}}$
\beq\label{4.8}
a^{-n}\sum_{p=0}^{n+2}C_p^{n+2}(-)^{n-p}W^{(3)
{\rm lead}}_{3p}=\hat{\cal W}_n^{(3){\rm lead}},
\ \ n\geq-2,
\eeq
where $\hat {\cal W}^{(3){\rm lead}}$ is given by (\ref{3.2}) with shifted
limits of the summation
\bea\label{4.9}
\sqrt3 \hat{\cal W}_n^{(3){\rm lead}}=
&\sum_{k+m\geq-n}
\left(k+{1\over3}\right)\left(m+{1\over3}\right)T_{3k+1}T_{3m+1}
{\partial\over\partial\tilde T_{3(k+m+n)+2}}\nn\\
+&\sum_{k+m\geq-n-1}
\left(k+{2\over3}\right)\left(m+{2\over3}\right)T_{3k+2}T_{3m+2}
{\partial\over\partial\tilde T_{3(k+m+n-1)+1}}.
\eea
The summation in (\ref{4.9}) includes the terms with times $T_m$,
$m<0$ defined by trivial continuing the equation (\ref{4.5}).

The proof of (\ref{4.8}) is based on the formulas (\ref{4.5}), (\ref{4.6})
and the following identity for the $\Gamma$-functions
\beq\label{3.3}
 \sum_{\gamma=0}^a{(-)^\gamma\Gamma(\gamma+b)\over \gamma!(a-\gamma)!
\Gamma(c+\gamma)}={\Gamma(b)\Gamma(c-b+a)\over \Gamma(a+1)\Gamma(c+a)
\Gamma(c-b)}.
\eeq

To eliminate the incorrect negative mode terms in (\ref{4.9}) we use
the partition function rescaled
by an exponential of a quadratic form of the auxiliarly times
$\tilde T_{3n+i}$ (compare with (\ref{a11}))
\beq\label{4.10}
\tilde\tau=\exp\left(-\sum_{m,n\geq0}A_{mn}\tilde T_{3m+1}
\tilde T_{3n+2} \right)\tau^{\rm red}_{N_1,N_2},
\eeq
where $\tau^{\rm red}_{N_1,N_2}$ is the partition function of the
conformal two-matrix model after the reduction (\ref{4.4})
and
\beq\label{4.11}
A_{nm}={\Gamma\left(n+{4\over3}\right)\Gamma\left(m+{5\over3}\right)\over
\Gamma\left({1\over3}\right)\Gamma\left({2\over3}\right)}
{(-)^{n+m}a^{-n-m-1}\over n!m!(n+m+1)(n+m+2)}.
\eeq

This rescaling of the partition function is equivalent
to the following transformation of the  generators $\hat{\cal W}^{(3)}_n$
\beq\label{4.12}
\tilde{\cal W}^{(3)}_n=\exp{\left(-\sum_{m,n\geq0}A_{mn}\tilde T_{3m+1}
\tilde T_{3n+2} \right)}\hat{\cal W}^{(3)}_n
\exp{\left(\sum_{m,n\geq0}A_{mn}\tilde T_{3m+1}
\tilde T_{3n+2} \right)},
\eeq
where now not only leading terms should be taken into account in
(\ref{4.12}).

Then direct calculations similar to the calculations
in \cite{MMMM91} show that the
changing of the times (\ref{4.5}), (\ref{4.6}) and the rescaling (\ref{4.10})
establish the relation like (\ref{4.8})
\beq\label{4.13} {1\over\tilde\tau}\tilde{\cal
W}_p^{(3)}\tilde\tau=
a^{-p}\sum_{n=0}^{p+2}C_{p+2}^n(-)^{p-n}
\frac{W^{(3)\ {\rm red}}_{3n}\tau^{\rm red}}{\tau^{\rm red}},\
\ p\geq -2,
\eeq
where the generators
$W^{(3)\ {\rm red}}_{3n}$ depend only on the times $t_{3n}^{(1,2)}$
and derivatives with respect to them, and the generators $\tilde{\cal
W}_p^{(3)}$ are exactly the generators of the $W^{(3)}$-symmetry
of the continuum model \cite{FKN91a}.

By definition, the generators $W^{(3)}_{n}$ annihilate the partition
function (\ref{21}) for $p=3$  while
$W^{(3)\ {\rm red}}_{3n}\tau^{\rm red}$
does not need to vanish, since
\beq\label{4.14}
0 = W^{(3)}_{3n}\tau\Bigm\vert_{t_{3k+i}=0;\ i=1,2} =
W^{(3)\ {\rm red}}_{3n}\tau^{\rm red} +
\ldots\ \ \ ,
\eeq
where $\ldots$ means terms similar to $
\sum_{m,k} {\partial ^3 \log \tau /
\partial t_{3m+1} \partial t_{3n-3(m+k)+1} \partial t_{3k-2}}$ with
all possible {\it correct} gradations which do not vanish under the reduction
(\ref{4.4}).
In analogy with the results of \cite{MMMM91},
these additional terms provides the eq.(\ref{4.13})
can be rewritten in the terms of the {\it cubic} root of the $\tau$-function
$\sqrt[3]{\tilde\tau}$ (see the sect.5.2 for detailed explanations)
\beq\label{4.15}
{{\cal W}^{(3)\ {\rm cont}}_n\sqrt[3]{\tilde\tau}\over\sqrt[3]{\tilde\tau}}
=a^{-n}\sum_{p=0}^{n+2}C^p_{n+2}(-)^{n-p}\left. {W^{(3)}_{3p}\tau
\over\tau}
\right\vert_{t_{3k+i}=0;\ i=1,2}\left( 1+  O(a) \right).
\eeq

We exclude here all simple but tedious calculations of general case,
since for our consideration of CMM below we will use a more efficient way to
deal with the changing of the time-variables $t \to T$
(which was
also proposed in \cite{MMMM91}), that is, a scalar field formalism.
For example, in the case of the one-\mm
the Kazakov change of the time variables (\ref{a5}), (\ref{a6}) can be
deduced from the following prescription. Let us consider the free
 scalar field with periodic boundary conditions ((\ref{d7}) for $p=2$)
\beq\label{a8}
\partial\varphi(u)=\sum_{k\geq0}g_ku^{2k-1}+
\sum_{k\geq1}{\partial\over\partial t_{2k}}u^{-2k-1},
\eeq
and analogous scalar field with antiperiodic boundary conditions (\ref{b3a}):
\beq\label{a9}
 \partial\Phi(z)=\sum_{k\geq0}\left(\left(k+{1\over2}\right)T_{2k+1}
z^{k-{1\over2}}+
{\partial\over\partial \tilde T_{2k+1}}z^{-k-{3\over2}}\right).
\eeq
Then the equation
\beq\label{a10}
{1\over\tilde\tau}\partial\Phi(z)\tilde\tau =
a {1\over\tau^{\rm red}}\partial\varphi(u)\tau^{\rm red},\quad
u^2=1+az
\eeq
generates the correct transformation rules
(\ref{a5}), (\ref{a6}) and gives rise to the expression (\ref{a12})
for $A_{nm}$.
Taking  the square of the both sides of the identity (\ref{a10}),
\bea\label{a100}
&{1\over\tilde\tau}{\cal T}(z)\tilde\tau={1\over \tau^{\rm red}}T(u)\tau^{\rm
 red},
\eea
one can obtain after simple  calculations that the
relation (\ref{a15}) is valid.

\section{The case of $W^{(3)}$ in scalar field formalism}

\subsection{Scalar field formalism}

In this section we would like to consider the case of $W^{(3)}$-algebra in
full details in the framework of the scalar field formalism described
at the end of the previous section. In this subsection we describe the
formalism, and, in the following ones, we
 apply it to Virasoro and $W^{(3)}$-algebra
respectively.

Thus, let us consider the set of scalar fields for the discrete two-\mm
\bea\label{4.16}
\partial\varphi^{(1)}(u)=\sum_{k\geq0}g_k^{(1)}u^{3k-1}+
\sum_{k\geq1}{\partial\over\partial t^{(2)}_{3k}}u^{-3k-1},\nn\\
\partial\varphi^{(2)}(u)=\sum_{k\geq0}g_k^{(2)}u^{3k-1}+
\sum_{k\geq1}{\partial\over\partial t^{(1)}_{3k}}u^{-3k-1},
\eea
and  the scalar fields of the continuum model
\cite{FKN91a}
\bea\label{4.17}
& \partial\Phi^{(1)}(z)=\sum_{k\geq0}\left(\left(k+{1\over3}\right)T_{3k+1}
z^{k-{2\over3}}+
{\partial\over\partial \tilde T_{3k+2}}z^{-k-{5\over3}}\right),\nn\
\\
& \partial\Phi^{(2)}(z)=\sum_{k\geq0}\left(\left(k+{2\over3}\right)T_{3k+2}
z^{k-{1\over3}}+
{\partial\over\partial \tilde T_{3k+1}}z^{-k-{4\over3}}\right).
\eea

Then all the relations (\ref{4.5}), (\ref{4.6}) and (\ref{4.11}) can be
encoded in the equations
\beq\label{4.18}
{1\over\tilde\tau}\partial\Phi^{(i)}(z)\tilde\tau =
a u^{i-2}{1\over\tau^{\rm red}}\partial\varphi^{(i)}(u)\tau^{\rm
red},~~\
u^3=1+az,\ i=1,2.
\eeq

Let us prove (\ref{4.18}) for the case of $i=2$ only. The $i=1$ case can be
treated analogously.
We have to check that the relation $u^3=1+az$, with using (\ref{4.5}) and
(\ref{4.6}), gives rise to the following equation
\bea\label{4.19}
&  a\sum_{k\geq0}g^{(2)}u^{3k-1}+a\sum_{k\geq1}{\partial\ln\tau\over\partial
t_{3k}^{(1)}}u^{-3k-1}=\nn\\
&  =\sum_{k\geq0}\left(k+{2\over3}\right)T_{3k+2}z^{k-{1\over3}}+
a\sum_{k\geq0}\left({\partial\ln\tau\over\partial
\tilde T_{3k+1}}z^{-k-{4\over3}}-\sum_{m\geq0}A_{mk}\tilde
T_{3m+2}z^{-k-{4\over3}}\right).
\eea

Let us start our check with the derivative terms. Using the obvious relations
\beq\label{4.20}
u^{3\alpha}=(1+az)^{\alpha}=\sum_{m\geq0}{\Gamma\left(\alpha+1
\right)\over m!\Gamma\left(\alpha+1-m\right)}(az)^{\alpha-m}
\eeq
and (\ref{4.6}), we obtain
\bea\label{4.21}
& \sum_{k\geq1}{\partial\over\partial
t_{3k}^{(1)}}u^{-3k-1}=\nn\\
& =\sum_{k\geq1}\sum_{f=0}^{k-1}\sum_{m\geq0}
{\Gamma\left(k+{1\over3}\right)\Gamma\left(-k+{2\over3}\right)
a^{f-m-k}\over(k-f-1)!\Gamma\left(f+{4\over3}\right)
m!\Gamma\left(-k-m+{2\over3}\right)}
{\partial\over\partial\tilde T_{3f+1}}z^{-m-k-{1\over3}}=\nn\\
& =\sum_{f\geq0}{a^f\over\Gamma\left(f+{4\over3}\right)}
{\partial\over\partial\tilde T_{3f+1}}
\sum_{k\geq0}\sum_{m\geq0}
{\pi\sin\left(\pi k+{4\pi\over3}\right)
a^{-m-k-1}\over(k-f)!m!\Gamma\left(-k-m-{1\over3}\right)}
z^{-m-k-{4\over3}}=\nn\\
& =\sum_{f\geq0}\sum_{\gamma\geq0}{a^{f-\gamma-1}
\pi\sin\left(\pi
\gamma+{4\pi\over3}\right)\over\Gamma\left(f+{4\over3}\right)
\Gamma\left(-\gamma-{1\over3}\right)}
{\partial\over\partial\tilde T_{3f+1}}
z^{-\gamma-{4\over3}}\delta_{f,\gamma}=
a^{-1}\sum_{f\geq0}
{\partial\over\partial\tilde T_{3f+1}}
z^{-f-{4\over3}},
\eea
where we replaced the variable of the summation $k\to \gamma=k+m$ and used the
identity
\beq\label{4.22}
\sum_{m\geq0}^{\gamma}{(-)^m  \over
m!(\gamma-f-m)!}=\delta_{f,\gamma}.
\eeq

The first sum in the $l.h.s.$ of (\ref{4.19}), after the substitution
$u^3=1+az$ and using (\ref{4.5}), falls into two different sums
\bea\label{4.23}
\sum_{k\geq0}kt_{3k}^{(2)}u^{3k-1}=&
\sum_{k\geq0} \sum_{n\geq k}\sum_{m=0}^{k}
{\Gamma\left(n+{5\over3}\right)(-)^{n-k}a^{k-m-n-1}\over
\Gamma\left(k-m+{2\over3}\right)(n-k)!m!}T_{3n+2}z^{k-m-{1\over3}}\nn\\
&+\sum_{k\geq0} \sum_{n\geq k}\sum_{m\geq k+1}
{\Gamma\left(n+{5\over3}\right)(-)^{n-k}a^{k-m-n-1}\over
\Gamma\left(k-m+{2\over3}\right)(n-k)!m!}T_{3n+2}z^{k-m-{1\over3}}.
\eea
The first sum in (\ref{4.23}) can be rewritten in the form
\bea\label{e13}
&\sum_{n\geq0} \sum_{k=0}^{n}\sum_{m\geq k}
{\Gamma\left(n+{5\over3}\right)(-)^{n-k}a^{k-m-n-1}\over
\Gamma\left(k-m+{2\over3}\right)(n-k)!m!}T_{3n+2}z^{k-m-{1\over3}}=\nn\\
=&\sum_{n\geq0} \sum_{\gamma=0}^{n}\sum_{k=\gamma}^{n}T_{3n+2}
{\Gamma\left(n+{5\over3}\right)\over\Gamma\left(\gamma+{2\over3}\right)}
(-)^{n}a^{\gamma-n-1}{(-)^k\over
(n-k)!(k-\gamma)!}z^{\gamma-{1\over3}}=\nn\\
=&\sum_{n\geq0} \sum_{\gamma=0}^{n} T_{3n+2}
{\Gamma\left(n+{5\over3}\right)\over\Gamma\left(\gamma+{2\over3}\right)}
a^{\gamma-n-1}z^{\gamma-{1\over3}}\delta_{n,\gamma}=
a^{-1}\sum_{n\geq0} \left(n+{2\over3}\right)T_{3n+2}
z^{n-{1\over3}},
\eea
where we
used the identity
(\ref{4.22}). After introducing the new variable of the
summation $\beta=m-k-1$,
changing the order of the summation and using the identity
\beq\label{4.25}
\sum_{k=0}^{n}{(-)^k\over(n-k)!(k+\beta+1)!}={1\over\beta!n!(\beta+n+1)},
\eeq
the second
sum in the $r.h.s.$ of (\ref{4.23}) can be rewritten in the form
\bea\label{4.26}
a^{-1}&\sum_{\beta\geq0} \sum_{n\geq0}
\left(\tilde T_{3n+2}+a{n\over n+{2\over3}}\tilde T_{3n-1}\right)
{\Gamma\left(n+{5\over3}\right)\over\Gamma\left(-\beta-{1\over3}\right)}
{(-)^{n}a^{-n-1-\beta}\over
(\beta)!n!(\beta+n+1)}
z^{-\beta-{4\over3}}=\nn\\
&=a^{-1}\sum_{\beta\geq0} \sum_{n\geq0}
\tilde T_{3n+2}
{\Gamma\left(n+{5\over3}\right)\over\Gamma\left(-\beta-{1\over3}\right)}
{(-)^{n}a^{-n-\beta-1}\over
(\beta)!n!(\beta+n+1)(\beta+n+2)}
z^{-\beta-{4\over3}}=\nn\\
&=-a^{-1}\sum_{\beta\geq0} \sum_{n\geq0}
A_{n\beta}\tilde T_{3n+2}
z^{-\beta-{4\over3}},
\eea
and from the last two lines of (\ref{4.26}) the expresion (\ref{4.11})
for the matrix
$A_{nm}$ follows.

Performing the similar calculations for the case of
$i=1$ in (\ref{4.18}), one can
find the same expression for the matrix $A_{mn}$. So,
we may conclude that
the changing of times (\ref{4.5}), (\ref{4.6}) results
in the equations (\ref{4.18}).

\subsection{Virasoro constraints of the two-\mm}

It was proposed in \cite{FKN91a} that the  continuum
``two-matrix'' model  {\em possesses}  the ${\cal W}^{(3)}$
and Virasoro symmetries,
the Virasoro generators ${\cal L}_n$ and the
generators of the ${\cal W}^{
(3)}$
algebra being constructed from the scalar fields (\ref{4.17})
in the following way
\beq\label{4.27}
{\cal T}(z)={1\over2}
{:}\partial\Phi^{(1)}(z)\partial\Phi^{(2)}(z){:}-{1\over9z^2}
=\sum{{\cal L}_n\over
z^{n+2}},
\eeq
\beq\label{4.28}
{\cal W}^{(3)}(z)={1\over3\sqrt3}
\left({:}\left(\partial\Phi^{(1)}(z)\right)^3{:}+
{:}\left(\partial\Phi^{(2)}(z)\right)^3{:}\right)
=\sum{{\cal W}_n^{(3)}\over
z^{n+3}}.
\eeq

We will prove  below that the equations (\ref{4.18}) yield the relation
between  the generators $\tilde{\cal L}_n$
and the corresponding generators $W_{3n}^{(2)\ {\rm red}}=
L_{3n}^{{\rm red}}$ associated with
the reduction (\ref{4.4}):
\beq\label{4.29}
{1\over\tilde\tau}\tilde{\cal L}_n\tilde\tau=
a^{-n}\sum_{p=0}^{n+1}C_p^{n+1}(-)^{n+1-p}
{L_{3p}^{\rm red}\tau^{\rm red}\over\tau^{\rm red}},\
\ \ n\geq-1,
\eeq
where Virasoro generators $L_{3n}^{\rm red}$ are defined by the  formula
(\ref{d3}), where only $t^{(i)}_{3n}$ and corresponding derivatives are
nonzero and the generators
\bea\label{4.30} \tilde{\cal
L}_{-1}=&\sum_{k\geq1}\left(\left(k+{1\over3}\right)T_{3k+1}
{\partial\over\partial \tilde T_{3k-2}}+
\left(k+{2\over3}\right)T_{3k+2}
{\partial\over\partial \tilde T_{3k-1}}\right)~+~{2\over9}T_1T_2,\nn\\
\tilde{\cal L}_{0}=&\sum_{k\geq0}\left(\left(k+{1\over3}\right)T_{
3k+1}
{\partial\over\partial \tilde T_{3k+1}}+
\left(k+{2\over3}\right)T_{3k+2}
{\partial\over\partial \tilde T_{3k+2}}\right), \nn\\
\tilde{\cal L}_n=&\sum_{k-m=-n}\left(\left(k+{1\over3}\right)T_{3k+1}
{\partial\over\partial \tilde T_{3m+1}}+
\left(k+{2\over3}\right)T_{3k+2}
{\partial\over\partial \tilde T_{3m+2}}\right)  \nn\\
&+\sum_{m+k=n-1}{\partial\over\partial \tilde T_{3k+2}}
{\partial\over\partial \tilde T_{3m+1}} +{(-)^n\over9a^n},\ \ \
n\geq1
\eea
differ from the continuum generators ${\cal
L}_n$ of the ref.\cite{FKN91a} by singular $c$-number terms. Indeed, again
$L_{3n}^{\rm red}$'s do not annihilate $\tau^{\rm red}$ exactly, but
these two effects cancel each other, provided eq.(\ref{4.29}) for $
\tilde\tau$
is rewritten in terms of the {\em cubic} root  of the $\tau$-function
$\sqrt[3]{\tilde\tau}$.
In other words, doing accurately the reduction procedure in the
Virasoro
constraints of the discrete two-\mm one can rewrite
(\ref{4.29}) in the form
\beq\label{4.31}
{1\over\sqrt[3]{\tilde\tau}}{\cal L}^{\rm cont}_n\sqrt[3]{\tilde\tau}=
a^{-n}\sum_{p=0}^{n+1}C_p^{n+1}(-)^{n+1-p}\left.{L_{3p}\tau\over\tau}
\left( 1+{\cal O}(a) \right)\right\vert_{t_{3k+i}=0;\  i=1,2},\
\ \ n\geq-1.
\eeq
Thus, we conclude that, indeed,
the continuum Virasoro constraints for the case of $p=3$ can be derived from
the corresponding Virasoro constraints of the discrete {\em conformal}
two-matrix model.

To prove (\ref{4.29})  let us calculate
\bea\label{4.32}
&{\Phi^{(1)}(z)\tilde\tau\over\tilde\tau}{\Phi^{(2)}(z)
\tilde\tau\over\tilde\tau}  =\sum_{n\geq0}z^{-n-2}
\left[\sum_{k\geq0}\left(k+{1\over3}\right)
T_{3k+1}{\partial\ln\tilde\tau\over\partial\tilde T_{3(k+n)+1}}\right.+\nn\\
&\left.\quad +
\left(k+{2\over3}\right)
T_{3k+2}{\partial\ln\tilde\tau\over\partial\tilde
T_{3(k+n)+2}}+\sum_{k=0}^{n-1}{\partial\ln\tilde\tau\over\partial\tilde
T_{3k+2}} {\partial\ln\tilde\tau\over\partial\tilde
T_{3(n-1-k)+2}}\right]\nn\\
&\quad +{1\over z}
\left[\sum_{k\geq1}\left(k+{1\over3}\right)
T_{3k+1}{\partial\ln\tilde\tau\over\partial\tilde T_{3(k-1)+1}}+
\left(k+{2\over3}\right)
T_{3k+2}{\partial\ln\tilde\tau\over\partial\tilde
T_{3(k-1)+2}}+{2\over9}T_1T_2\right],
\eea
where in (\ref{4.32}) only singular at $z\to\infty$ terms are taken
into account.

Using at $n\geq1$ the identity
\beq\label{4.33}
\sum_{k+m=n-1\atop k,m\geq0}A_{km} =\sum_{k+m=n-1\atop k,m\geq0}
{\Gamma\left(k+{5\over3}\right)\Gamma\left(m+{4\over3}\right)\over
k!m!\Gamma\left({2\over3}\right)\Gamma\left({1\over3}\right)}
{(-)^{n-1}a^{-n}\over n(n+1)}={(-)^{n-1}\over9a^n},
\eeq
the formula (\ref{4.32}) can be rewritten in the compact form
\beq\label{4.34}
{\Phi^{(1)}(z)\tilde\tau\over\tilde\tau}{\Phi^{(2)}(z)
\tilde\tau\over\tilde\tau}
=\sum_{n\geq-1}z^{-n-2}\left[{\tilde{\cal L}_n\tilde\tau\over\tilde\tau}
-\sum_{k+m=n-1}{\partial^2\ln\tau\over\partial\tilde T_{3k+2}
\partial\tilde T_{3m+1}}\right].
\eeq

Using the fact that
the generators $L_{3n}^{\rm red}$ of the discrete Virasoro algebra
can be
obtained from the formula
\beq\label{4.35}
\sum_{n}u^{-3n-2}L_{3n}=
{1\over2}{:}\partial\varphi^{(1)}(u)\partial\varphi^{(2)}(u){:}\ \ ,
\eeq
one can easily show that
\beq\label{4.36}
{a^2\over u} {\partial\varphi^{(1)}(u)\tau\over\tau}
{\partial\varphi^{(2)}(u)\tau\over\tau}=a^2\sum_{n\geq0}u^{-3(n+1)}
\left[{L_{3n}\tau\over\tau}-\sum_{m+k=n\atop m,k\geq1}
{\partial^2\ln\tau\over\partial t^{(2)}_{3k}t^{(1)}_{3m}}\right],
\eeq
where again only the terms singular at $u\to\infty$ are taken into
account. Now using the equations (\ref{4.18}) we can
conclude that
\bea\label{4.37}
&\sum_{n\geq-1}z^{-n-2}\left[{\tilde{\cal L}_n\tilde\tau\over\tilde \tau}
-\sum_{k+m=n-1}{\partial^2\ln\tau\over\partial\tilde T_{3k+2}
\partial\tilde T_{3m+1}}\right]=\nn\\
&\quad=a^2\sum_{n\geq0}u^{-3(n+1)}
\left[{L_{3n}\tau\over\tau}-\sum_{m+k=n\atop m,k\geq1}
{\partial^2\ln\tau\over\partial t^{(2)}_{3k}t^{(1)}_{3m}}\right].
\eea
It follows now from (\ref{4.20}) that
\beq\label{4.38}
{\tilde {\cal L}_n\tilde\tau\over\tilde\tau}
-\sum_{k+m=n-1\atop k,m\geq0}{\partial^2\ln\tau\over\partial\tilde
T_{3k+2} \partial\tilde T_{3m+1}}=
a^{-n}\sum_{p=0}^{n+1}C^p_{n+1}(-)^{n+1-p}
\left[{L_{3p}\tau\over\tau}-\sum_{m+k=p\atop m,k\geq1}
{\partial^2\ln\tau\over\partial t^{(2)}_{3k}t^{(1)}_{3m}}\right].
\eeq
If we take into account that
\bea\label{4.39}
&a^{-n}\sum_{p=0}^{n+1}(-1)^{n+1-p}C_p^{n+1}\sum_{m+k=p\atop
m,k\geq1}
{\partial^2\over\partial t^{(2)}_{3k}t^{(1)}_{3m}}=\nn\\
&=a^{-n}(-1)^{n+1}\sum_{f\geq0}\sum_{g\geq0}{\partial^2\over\partial
\tilde T_{3f+1}\partial\tilde T_{3g+2}}
{a^{f+g+1}\over\Gamma\left(f+{4\over3}\right)\Gamma\left(g+{5\over3}\right) }
\nn\\
&\times   \sum_{p=2}^{n+1}\sum_{m=1}^{p-1}(-1)^pC_{n+1}^p
{\Gamma\left(m+{1\over3}\right)\Gamma\left(p-m+{2\over3}\right)\over
(m-f-1)!(p-m-g-1)!}=\sum_{f+g=n-1\atop f,g\geq 0}
{\partial^2\over\partial\tilde T_{3f+1}\partial\tilde T_{3g+1}},
\eea
we may conclude that (\ref{4.29}) is correct.

\subsection{Connection between ${\cal W}^{(3)
}_n$ and $W^{(3)}_{3n}$}

Now we are going to repeat the procedure of the previous subsection
for the $W^{(3)}$-algebra generators.
The connection between the
discrete and the continuum cases was formulated in the
section 4.1 and based on the
formula (\ref{4.13}) which can be easily proved by means of the
equations (\ref{4.18}). Indeed, it is easy to see that the generators
 $W_n^{(3)\ {\rm red}}$ of the reduced discrete model
can be rewritten using the scalar fields (\ref{4.16}) as
\beq\label{4.40}
\sum_{n}u^{-3n-3} W^{(3)}_{3n}={1\over 3\sqrt3}
\left[{1\over
u^3}{:}\left(\partial\varphi^{(1)}(u)\right)^3{:}
+{:}\left(\partial\varphi^{(2)}(u)\right)^3{:}\right]\ \ .
\eeq
Then
the relation between the generators of the $W^{(3)}$ symmetry
for the discrete  model and the generators $\tilde{\cal W}_n^{(3)}$
is
\beq\label{4.41} {1\over\tilde\tau}\tilde{\cal
W}_p^{(3)}\tilde\tau=
a^{-p}\sum_{n=0}^{p+2}C_{p+2}^n(-)^{p-n}
\frac{W^{(3)\ {\rm red}}_{3n}\tau^{\rm red}}{\tau^{\rm red}},
\ p\geq -2
\eeq
and follows from the identity
\beq\label{4.42}
\left({\partial\Phi^{(1)}(z)\tilde\tau\over\tilde\tau}\right)^3
+\left({\partial\Phi^{(2)}(z)\tilde\tau\over\tilde\tau}\right)^3=
a^3\left[{1\over
u^3}\left({\partial\varphi^{(1)}(u)\tau^{\rm red}\over\tau^{\rm red}}\right)^3
+\left({\partial\varphi^{(2)}(u)\tau^{\rm red}\over\tau^{\rm red}}
\right)^3\right].
\eeq

The generators $\tilde{\cal W}_n^{(3)}$ $n\geq-2$ are defined by the formulas
(\ref{4.28})   and the generators $\tilde{\cal W}_{-2}^{(3)}$  and
$\tilde {\cal W}_{-1}^{(3)}$ (but
not the $\tilde{\cal W}_0^{(3)}$) have additional
terms, cubic in times:
\beq\label{e31}
{\cal W}_{-1}^{(3)}={1\over27}T_1^3+\cdots\ \ \hbox{and}\ \
{\cal W}_{-2}^{(3)}={8\over27}T_2^3+{4\over9}T_1^2T_4+\cdots\ .
\eeq
 To obtain (\ref{4.41}) from (\ref{4.42})
one need the set of the combinatorial identities
\bea\label{4.43}
a^{-n}\sum_{f=0}^{n+2}C_{n+2}^f (-)^{n-f}\sum_{k+m+p=f}
{\partial^3
\over\partial t^{(i)}_{3k}\partial t^{(i)}_{3m}\partial t^{(i)}_{3p}}
=   \sum_{\alpha+\beta+\gamma=n-1}{\partial^3\over
\partial T_{3\alpha+i}\partial T_{3\beta+i}\partial T_{3\gamma+i}},
\nn
\eea
\bea\label{4.44}
&a^{-n}\sum_{f=0}^{n+2}C_{n+2}^f (-)^{n-f}\sum_{m,p\geq1}
\left(g^{(1)}_{m+p-n+1}+{\partial\ln\tau\over\partial t^{(2)}
_{n-1-m-p}}\right){\partial^2\over\partial t^{(2)}_{3m}
\partial t^{(2)}_{3p}}=\nn\\
=&
\sum_{\beta+\gamma-\alpha=n-1}\left(\alpha+{1\over3}\right)T_{3\alpha+1}
{\partial^2\over\partial\tilde
T_{3\beta+2}\partial\tilde T_{3\gamma+2}}
+
\sum_{\beta+\gamma+\alpha=n-2}{\partial\ln\tilde\tau\over\partial\tilde
T_{3\alpha+2}}
{\partial^2\over\partial\tilde
T_{3\beta+2}\partial\tilde T_{3\gamma+2}},
\nn
\eea
\bea\label{4.45}
&a^{-n}\sum_{f=0}^{n+2}C_{n+2}^f (-)^{n-f}\sum_{m,p\geq1}
\left(g^{(2)}_{m+p-n}+{\partial\ln\tau\over\partial t^{(1)}
_{n-m-p}}\right){\partial^2\over\partial t^{(1)}_{3m}
\partial t^{(1)}_{3p}}=\nn\\
=&
\sum_{\beta+\gamma-\alpha=n}\left(\alpha+{2\over3}\right)T_{3\alpha+2}
{\partial^2\over\partial\tilde
T_{3\beta+1}\partial\tilde T_{3\gamma+1}}
+
\sum_{\beta+\gamma+\alpha=n-1}{\partial\ln\tilde\tau\over\partial\tilde
T_{3\alpha+1}}
{\partial^2\over\partial\tilde
T_{3\beta+1}\partial\tilde T_{3\gamma+1}}.
\nn\eea
which can be proved by the direct calculation using the formulas (\ref{4.5}),
(\ref{4.6}) and identities for the $\Gamma$-functions similar to
\beq
\sum_{m=0}^d{\Gamma(m+b)\Gamma(c-m)\over m!(a-m)!}=
{\Gamma(c-d)\Gamma(b)\Gamma(b+c)\over d!\Gamma(b-c-d)}.\nn
\eeq

\section{General case}

\subsection{The continuum limit of $W^{(n)}$-algebra}

It is easy to generalize the formailsm of the scalar field
described in the sect.4.2-4.3 to the general conformal multi-\mms
using the scalar fields
with ${\bf Z}_p$-twisted boundary conditions.
Let us introduce $p-1$ sets of the discrete times
$t^{(i)}_{k}$, $i=1,2,\ldots,p-1$ and $k=0,1,\ldots$ for the discrete
$(p-1)$-\mm and consider the reduction
\beq\label{5.1}
t^{(i)}_{pk+j}=0,\ i,j=1,2,\ldots,p-1,\ k=0,1,\ldots\ \ .
\eeq
One can choose  the "discrete" and the "continuum"
 scalar fields in the form
\beq\label{5.2}
\partial\varphi^{(i)}(u)=\sum_{k\geq0}g_k^{(i)}u^{pk-1}+
\sum_{k\geq1}{\partial\over\partial t^{(p-i)}_{pk}}u^{-pk-1},
\quad g_k^{(i)}=kt^{(i)}_{pk}, \quad g_0^{(i)}=N_i\ ,
\eeq
\beq\label{5.3}
\partial\Phi^{(i)}(z)=\sum_{k\geq0}\left\{\left(k+{i\over p}\right)T_{pk+i}
z^{k-{p-i\over p}}+
{\partial\over\partial \tilde T_{pk+p-i}}z^{-k-{2p-i\over  p}}\right\},
\eeq
\beq\label{5.4}
  T_{pk+i}=\tilde T_{pk+i}+a{k\over k+i/p}\tilde T_{p(k-1)+i},\
i=1,2,\ldots,p-1 .
\eeq

Then the equations
\beq\label{5.5}
a u^{i-p+1}{1\over\tau^{\rm red}}\partial\varphi^{(i)}(u)\tau^{\rm red}=
{1\over\tilde\tau}\partial\Phi^{(i)}(z)\tilde\tau,\
u^p=1+az,\ i=1,2,\ldots,p-1,
\eeq
generate Kazakov-like change of the time variables
\beq\label{5.6}
g^{(i)}_m=\sum_{n\geq m}{(-)^{n-m}\Gamma\left(n+1+{i\over p}\right)
a^{-n-{i\over p}}\over(n-m)!\Gamma\left(m+{i\over p}\right)}T_{pn+i},\
i=1,2,\ldots,p-1
\eeq
\beq\label{5.7}
{\partial\over\partial t^{(i)}_{pk}}=\sum_{n=0
}^{k-1}{\Gamma\left(k+{i\over p}\right)
a^{n+{i\over p}}\over(k-n-1)!\Gamma\left(n+1+{i\over p}\right)}{\partial
\over\partial \tilde T_{pn+i}},\ i=1,2,\ldots,p-1.
\eeq

Then the consideration similar to
those used in the previous
subsections  shows that the relation
between the tilded continuum generators $\tilde{\cal W}^{(i)}_n$,
$i=2,\ldots,p$  of the ${\cal W}$-symmetry and the reduced generators of
the discrete $W$-symmetry $W_{pk}^{(i){\rm red}}$ is as follows
\beq\label{5.8}
{1\over\tilde\tau}\tilde{\cal W}_n^{(i)}\tilde\tau=
a^{-n}\sum_{s=0}^{n+i-1} C^s_{n+i-1}(-)^{n+i-1-s}
\frac{W^{(i){\rm red}}_{ps}\tau^{\rm red}}
{\tau^{\rm red}},\ \ \ \  n\geq -i+1,
\eeq
where the rescaled $\tau$-function is defined
\beq\label{5.9}
\tilde\tau=\exp\left(-{1\over2}\sum_{i=1}^{p-1}\sum_{m,n\geq0}A^{(i)}_{mn}
\tilde T_{pm+i}
\tilde T_{pn+p-i} \right)\tau^{\rm red}\{t_{pk}\}
\eeq
and the matrices $A^{(i)}_{nm}$ are determined by
\beq\label{5.10}
A^{(i)}_{nm}={\Gamma\left(n+{p+i\over p}\right)\Gamma\left(m+{2p-i\over p}
\right)\over
\Gamma\left({i\over p}\right)\Gamma\left({p-i\over p}\right)}
{(-)^{n+m}a^{-n-m-1}\over n!m!(n+m+1)(n+m+2)},\  i=1,2,\ldots,p-1.
\eeq

The proof of the equivalence of (\ref{5.5}) to (\ref{5.6}), (\ref{5.7}),
(\ref{5.9}) and (\ref{5.10}) is the same as in the sect.4.2-4.3 and
we omit it here. To prove the equations (\ref{5.8}), we need the exact
expressions how the generators
$\tilde{\cal W}^{(i)}_n$   and $W_{k}^{(i){\rm red}}$
are connected with corresponding scalar fields
(\ref{5.2}), (\ref{5.3}).

Performing the proper reduction procedure (\ref{5.1}), which eliminates all
time-variables excepting those of the form $t_{pk}^{(i)}$
(i.e. leaves the $1/p$ fraction
of the entire quantity of variables) we can obtain the relation
\beq\label{5.11}
{1\over\sqrt[p]{\tilde\tau}}{\cal W}_n^{(i)}\sqrt[p]{\tilde\tau}=
a^{-n}\sum_{s=0}^{n+i-1}C^s_{n+i-1}(-)^{n+i-1-s}\left.{W^{(i)}_{ps}
\tau\over\tau}\right\vert_{t_{pk}^{(i)}\neq 0\ {\rm only}}
\left( 1+ O(a) \right),\ \ \ n\geq-i+1,
\eeq
where ${\cal W}_n^{(i)}$'s are  the ${\cal W}$-generators of the paper
\cite{FKN91a}.
Thus, we proved the $W$-invariance of the partition function of the
continuum $p-1$-\mm and found the explicit connection of its
partition function with
the corresponding partition function of the discrete $(p-1)$-matrix model.

\subsection{On the reduction of the partition function}

To conclude this section we would like to discuss the problem of reduction
(\ref{4.4}) and (\ref{5.1}) of the partition function in detail,
with accuracy up to (non-leading) $c$-number contributions and only
after the continuum limit is taken. More
precisely, we reformulate the  condition of a proper reduction in the continuum
limit in order to reduce it to more explicit formulas which can be
immediately checked. As a by-product of our consideration we obtain some
restrictions on the integration contour in the partition function (\ref{21}).

To get some insight, let us consider the simplest case of the Virasoro
constrained Hermitian one-\mm \cite{MMMM91}.
Before the reduction the Virasoro operators read as in
(\ref{b5}). Then their action on $\log \tau$ can be rewritten as
\beq\label{f1}
\left[ \sum_k kt_k {\partial \log \tau \over \partial t_{k+n}} +
\sum_m {\partial ^2 \log \tau \over \partial t_m \partial t_{n-m}}\right] +
\sum_m \left[{\partial \log \tau \over \partial t_m}
{\partial \log \tau \over \partial t_{n-m}}\right] = 0.
\eeq

After the reduction, we obtain
\bea\label{f2}
&\left[ \sum_k 2kt_{2k} {\partial \log \tau ^{\rm red}\over \partial t_{2k+2n}}
+
\sum_m {\partial ^2 \log \tau ^{\rm red} \over
\partial t_{2m} \partial t_{2n-2m}} +
\sum_m {\partial ^2 \log \tau ^{\rm red} \over
\partial t_{2m+1} \partial t_{2n-2m-1}} \right] + \nn\\
&+ \sum_m \left[{\partial \log \tau ^{\rm red} \over \partial t_{2m}}
{\partial \log \tau ^{\rm red} \over \partial t_{2n-2m}}\right] = 0
\eea
under the condition
\beq\label{f3}
\left.{\partial \log \tau ^{\rm red} \over \partial t_{\rm odd}}\right|_{t_{\rm
odd}=0} = 0.
\eeq

The last formula is a direct consequence of the ``Schwinger-Dyson'' equation
induced by the transformation of the reflection $M \to -M$ in (\ref{b4}).
Indeed, due to the invariance of the integration measure under this
transformation one can conclude that the partition function (\ref{b4}) depends
only on a quadratic form of odd times.

Thus, the second derivatives of $\log \tau$ over odd times do not vanish, and
are conjectured to satisfy the relation
\beq\label{f4}
\sum_m {\partial ^2 \log \tau ^{\rm red} \over
\partial t_{2m} \partial t_{2n-2m}} \sim
\sum_m {\partial ^2 \log \tau ^{\rm red} \over
\partial t_{2m+1} \partial t_{2n-2m-1}},
\eeq
where the sign "$\sim$" implies that this relation should be correct only
{\it after} taking the continuum limit.
In this case one obtains the final result (cf. (\ref{a16}))
\beq\label{f5}
\left[ \sum_k kt_{2k} {\partial \log \sqrt{\tau ^{\rm red}}
\over \partial t_{2k+2n}} +
\sum_m {\partial ^2 \log \sqrt{\tau ^{\rm red}} \over \partial t_{2m}
\partial t_{2n-2m}}\right] = 0.
\eeq

Thus, it remains to check the correctness of the relation (\ref{f4}).
To do this,
one should use the manifest equations of integrable (Toda chain) hierarchy,
and after direct but tedious calulations \cite{MMMM91} one obtains the result
different from the relation (\ref{f4}) by $c$-number terms which are singular
in the limit $a\to 0$ and just cancell corresponding items in (\ref{b6})
(this is certainly correct only after taking the continuum limit).

All this (rather rough) consideration can be easily generalized to the
$p$-\mm case.
In this case one should try to use all ${\cal W}_n^{(i)}$-constraints with
$2\le i\le p$. Thus, the second derivatives should be replaced by higher
order derivatives, and one obtain a series of equations like (\ref{f1}). It is
the matter of trivial calculation to check that these equations really give
rise to the
proper constraints satisfied by $\sqrt[p]{\tau}$ (cf. (\ref{4.15})
 and (\ref{5.11}))
provided by the two sets of the relations like (\ref{f3}) and (\ref{f4}) .

Namely, the analog of the relation (\ref{f3}) in the $p$-\mm case is the
cancellation of all derivatives with incorrect gradation, i.e. with the
gradation non-equal to zero by modulo $p$. The other relation (\ref{f4})
should be replaced now by the conditions of the equality (in the continuum
limit) of {\it all} possible terms with the same {\it correct}
gradation. In the simplest case of $p=3$ these are
\bea\label{f6}
&\sum_m {\partial ^2 \log \tau ^{\rm red} \over
\partial t_{3m+1} \partial t_{3n-3m-1}} \sim
\sum_m {\partial ^2 \log \tau ^{\rm red} \over
\partial t_{3m} \partial t_{3n-3m}},\nn\\
&\sum_{m,k} {\partial ^3 \log \tau ^{\rm red} \over
\partial t_{3m+1} \partial t_{3n-3(m+k)+1} \partial t_{3k-2}} \sim
\sum_{m,k} {\partial ^3 \log \tau ^{\rm red} \over
\partial t_{3m+2} \partial t_{3n-3(m+k)+2} \partial t_{3k-4}} \sim\nn\\
&\sim \sum_{m,k} {\partial ^3 \log \tau ^{\rm red} \over
\partial t_{3m+2} \partial t_{3n-3(m+k)+1} \partial t_{3k}} \sim
\sum_{m,k} {\partial ^3 \log \tau ^{\rm red} \over
\partial t_{3m} \partial t_{3n-3(m+k)} \partial t_{3k}}.
\eea

Again, this second condition is correct modulo some singular in the limit of
$a\to 0$ terms, which appear only in the case of even $p$. Unfortunately,
we do not know the way to prove this statement without using the integrable
equations, what is very hard to proceed in the case of higher $p$.

On the other hand, the cancellation of derivatives with
incorrect gradation can be trivially derived from the ``Schwinger-Dyson''
equations given rise by the transformations
$M\to \exp\left\{{2\pi ki\over p}\right\}M$ ($0<k<p$) of the integration
variable in the
corresponding matrix integral, the integration measure being assumed to be
invariant. In its turn, it implies that the integration contour, instead of
real line, should be chosen as a set of rays beginning in the origin
of the co-ordinate system with the angles between them being integer times
$\displaystyle{2\pi \over p}$
\footnote{Simultaneously it means that the size of all matrices
in (\ref{21a}) should be specially adjusted, say, to be times of $p$.}. This
rather fancy choice of the integration contour
is certainly necessary to preserve ${\bf Z}_p$-invariance of $p$-\mm system.

\section{Conclusion}

To conlude, we have demonstrated that the partition function of
a new class of multi-matrix models (CMM)
proposed in \cite{MMM91} can be described by the same constraint algebras as
in the case of the standard multi-matrix models \cite{BDKS90,CGM90,GM90c}.
It was conjectured in
\cite{FKN91a} that it defines the partition function in full. Therefore, one
can conclude that the partition function of CMM lies in the same universality
class (i.e. have the same (double scaling) continuum limit) as the standard
models. Certainly, there is an essential difference in these models both in
their formulation at the discrete level and in
the procedure of taking the continuum limit. Indeed, in contrast to the
standard
multi-matrix models, CMM's satisfy $W$-constraint algebra
 already at the discrete
level.
But as a price, it leads to very complicated (though still
of the nearest neighbours)
interaction of matrices in the matrix integral. From the other hand,
the correct reduction procedure corresponds to taking the square root of the
partition function in the
standard case, in contrast to the root of the
$p$-th degree, which should be involved for the proper reduction
 of the conformal $(p+1)$-matrix
model.

Now we would like to emphasize that the double scaling limit of the standard
multi-matrix models was never investigated honestly in the whole space of
possible coupling constants ($i.e.$ not nearby a critical point). The problem
certainly is that it is very difficult to describe any matrix model
in the whole space without any explicit constraint algebra imposed on
the partition function. Therefore, the continuum limit for multi-matrix models
proposed in \cite{FKN91a} should be considered rather as a definition
than as the theorem.

In fact, we know another case when multi-matrix model possesses \wi. It is
the partition function described in \cite{MMM92b}, whose continuum limit was
investigated in \cite{GvN91}. The arguments mentioned above suggest that it
should be possible to propose adequate procedure of taking the continuum
limit for the $W$-constraints in this case as well, but it is not done up to
now.

At last, we would like to stress some points which are to be understood
better.

1) In the paper we have proved the correspondence
between discrete and continuum constraint
algebras at the level of leading terms (see the sect.3.2), but the whole proof
taking into account all corrections is still unknown. It is not a problem to
do this in any concrete case, and we have demonstrated it in the case of
$W^{(3)}$-constraint algebra, but it is very difficult to deal with the
general case. One of the main reasons is that it is difficult to write down the
general expression for the generators of $W^{(n)}$ algebra (see, for example,
\cite{FKN91b}).

Nevertheless, the main statement should be correct. We think that
there should be more elegant and efficient way of doing the continuum limit,
and that to find out this one should understand deeper the role which is played
by the
scalar fields with different boundary conditions.

2) We do not know how it is possible
to prove the condition of proper reduction like (\ref{f6}). It was done
explicitly only
in the one-matrix model case (the relation (\ref{f4})), where we used
the equations of the integrable
hierarchy (Toda chain hierarchy) in the manifest way and after rather tedious
calculations proved the statement \cite{MMMM91}. But even in the case of
two-matrix model the integrable equations are very complicated (maybe, we
do not understand them completely) and it is absolutely unclear how one should
solve this problem in the general case. Again, we think it might be done in
a simpler way.

The solution to all these problems seems to be important for a better
understanding of the role of the double scaling continuum limit.
We hope to return to all these problems elsewhere.

\section{Acknowledgments}

We are grateful for stimulating discussions to S.Kharchev, A.Marshakov and
A.Morozov.

\section{Appendix}

In this appendix we consider Kazakov-like change of the time
variables in the general case of the conformal $p-1$-\mm as well as the
relation between the discrete and the continuum $W$-algebras which can be
easily extracted from the analyzes of the leading terms only (see the
sect.3.2).
We can use again the fact that the constraint algebra of
CMM, by construction,
 strongly resembles the structure of the corresponding continuum
constraint algebra. This assertion means that the generators
$W^{(p)}_{ps}$'s of the discrete algebra, after the reduction
(\ref{5.1}), depend on the scalar fields (\ref{5.2}) through
\beq\label{a.1}
\sum _{s\in{\bf Z}}{W^{(p)}_{ps}\over u^{ps+p}}\sim
\sum_{i=1}^{p-1}{:}\left(\partial\varphi^{(i)}(u)\right)^p{:}\ u^{p(i-p+1)},
\eeq
where the sign "$\sim$" means that we omit some numerical
  coefficient. Then it is
obvious from (\ref{a.1}) that the generators $W^{(p)}_{ps}$'s are
 given by the sum
 (with proper binomial coefficients) of the terms
\beq\label{a.2}
\left.\prod_{a=1}^j\sum_{k_a\geq0}g^{(i)}_{k_a}
\prod_{b=1}^{p-j}\sum_{m_b\geq1}{\partial\over\partial
t^{(p-i)}_{pm_b}}\right|_{\sum\limits_{b=1}^{p-j}m_b=\sum\limits
_{a=1}^jk_a+i-p+s+1}\ \ ,
\eeq
where $j=0,1,\ldots,p-1$, $i=1,2,\ldots,p-1$.

Now we are going to prove that the Kazakov change of the time variables
(\ref{5.6}), (\ref{5.7}) results in the relation
\bea\label{a.3}
&a^{-n}\sum_{s=0}^{n+p-1}C^s_{n+p-1}(-)^{n+p-1-s}
\left.\prod_{a=1}^j\sum_{k_a\geq0}g^{(i)}_{k_a}
\prod_{b=1}^{p-j}\sum_{m_b\geq1}{\partial\over\partial
t^{(p-i)}_{pm_b}}\right|_{\sum m_b=\sum k_a+i-p+s+1}\nn\\
&\quad=\left.\prod_{a=1}^j\prod_{b=1}^{p-j}
\sum_{n_a\atop\sum n_a\geq -n-i+p-j}
\sum_{f_b\geq0}\left(n_a+{i\over p}\right)T_{pn_a+i}
{\partial\over\partial
\tilde T_{pf_b+p-i}}\right|_{\sum f_b=\sum n_a+n+i-p+j},
\eea
where the sums over $a$ and $b$ in (\ref{a.3}) run in the same limits as
in (\ref{a.2}) and again in the sums
over $n_a$ some incorrect shifts
appear which should be removed by a proper rescaling of the
reduced partition function of CMM.  The expression
(\ref{a.3}) reproduces the
connection (\ref{5.8}).

The equation (\ref{a.3}) is based on the following identity
for the $\Gamma$-functions
\bea\label{a.4}
&\sum_{s=0}^{n+p-1}C^s_{n+p-1}(-)^{n+p-1-s}
\prod_{a=1}^j\sum_{k_a=0}^{n_a}{(-)^{n_a-k_a}\Gamma\left(n_a+{i\over p}
\right)\over (n_a-k_a)!\Gamma\left(k_a+{i\over p}\right)}\nn\\
&\left.\times \prod_{b=1}^{p-j}\sum_{m_b=f_b+1}^{\infty}
{\Gamma\left(m_b+{p-i\over p}\right)\over
(m_b-f_b-1)!\Gamma\left(f_b+{2p-i\over p}\right)}
\right|_{\sum m_b=\sum k_a+i-p+s+1}=\nn\\
&=\left\{\begin{array}{l}
1\quad{\rm if}\quad \sum f_b=\sum n_a+n+i-p+j\\
0\quad{\rm otherwise}, \end{array}\right.
\eea
which will be proved in several steps.

First, we note that the product  of the sums  over $b$
in the second line of (\ref{a.4})
can be easily simplified using the identity
\beq\label{a.5}
 \sum_{\gamma=0}^a{1\over \gamma!(a-\gamma)!
\Gamma(b-\gamma)\Gamma(c+\gamma)}={\Gamma(c+b+a-1)\over
\Gamma(a+1)\Gamma(b)\Gamma(c+a)\Gamma(c+b-1)},
\eeq
and the product  is equal to
\beq\label{a.6}
{\Gamma\left(\sum k_a+i-2p+s+1+j+(p-j){2p-i\over p}\right)\over
(\sum k_a-\sum f_b+i-2p+j+s+1)!
\Gamma\left(\sum f_b+(p-j){2p-i\over p}\right)}.
\eeq
Second, using the identity (\ref{3.3}) we perform the sum over $s$ in
(\ref{a.4}) to obtain
\bea\label{a.7}
&\prod_{a=1}^j\sum_{k_a=0}^{n_a}{(-)^{n_a-k_a}\Gamma\left(n_a+{i\over p}
\right)\over (n_a-k_a)!\Gamma\left(k_a+{i\over p}\right)}\nn\\
&\quad \times
{\Gamma\left(\sum k_a+i-p+1+j+(p-j){p-i\over p}\right)\over
\left(\sum k_a-\sum f_b-p+j+i+n\right)!
\Gamma\left(\sum f_b-n-p+1+(p-j){2p-i\over p}\right)}\ .
\eea
At last,  using the identities (\ref{a.5}) $j-1$ times and
(\ref{4.22}), we conclude that (\ref{a.4}) is correct.

\end{document}